# An Edge Computing Empowered Radio Access Network With UAV-Mounted FSO Fronthaul and Backhaul: Key Challenges and Approaches

Yanjie Dong, *Student Member, IEEE*, Md. Zoheb Hassan, *Student Member, IEEE*,
Julian Cheng, *Senior Member, IEEE*, Md. Jahangir Hossain, *Member, IEEE*,
and Victor C. M. Leung, *Fellow, IEEE*

## Abstract

One promising approach to address the supply-demand mismatch between the terrestrial infrastructure and the temporary and/or unexpected traffic demands is to leverage the unmanned aerial vehicle (UAV) technologies. Motivated by the recent advancement of UAV technologies and retromodulator based free space optical communication, we propose a novel edge-computing empowered radio access network architecture where the fronthaul and backhaul links are mounted on the UAVs for rapid event response and flexible deployment. The implementation of hardware and networking technologies for the proposed architecture are investigated. Due to the limited payload and endurance as well as the high mobility of UAVs, research challenges related to the communication resource management and recent research progress are reported.

Yanjie Dong, Md. Zoheb Hassan, and Victor C. M. Leung are with the Department of Electrical and Computer Engineering, The University of British Columbia, Vancouver, BC V6T 1Z4, Canada (e-mail: {ydong16, mdzoheb, vleung}@ece.ubc.ca).

Julian Cheng and Md. Jahangir Hossain are with the School of Engineering, The University of British Columbia, Kelowna, BC V1V 1V7, Canada (e-mail: {julian.cheng, jahangir.hossain}@ubc.ca).

## I. INTRODUCTION

Owing to the maturity of the unmanned aerial vehicle (UAV) technologies, various applications of UAVs have been proposed [1]–[3]. As an important use case, the UAV-assisted wireless communication has attracted much research attention [4]–[6] due to inherent advantages of UAVs, such as fast deployment, flexible configuration and possibility of having better channel conditions. When the terrestrial infrastructure fails to satisfy temporary and/or unexpected traffic demands, which is also referred to as supply-demand mismatch, the UAV-mounted infrastructure becomes an efficient alternate to meet the temporary and/or unexpected demands. The potential application scenarios can be a highway with traffic congestion, where traffic demands at a specific segment of the highway increase unexpectedly due to the accumulated user equipments (UEs). Deploying extra terrestrial infrastructure is not a timely solution in this case. For rapid event response, the UAV-mounted infrastructure can be deployed immediately by launching several specialized UAVs [2]. Another important application scenario is the rapid deployment of infrastructure in a region after a catastrophic event. Recovery of connection between the post-disaster region and the core network may require several weeks. The communication links during the recovery period can rely on satellites and UAVs. However, the UAV-mounted infrastructure are preferred over satellite infrastructure due to flexible deployment, low capital expenditure, and low transmission delay [2]. Recently, Qualcomm and AT&T have announced UAV connectivity trials with the objective of better serving emerging IoT demands in logistics, search and rescue, and inspection sectors.

Moreover, the future networks are envisioned to be highly user-centric, and the user demand information are generally in large quantity due to an exponential increase of terminals in the mobile networks. One promising solution to incorporate the user demands and rapid event response in the communication resource management (CRM) is via the edge computing technology [7]. Evolutionized from the cloud computing networks, the edge computing empowered radio access networks (EC-RANs) push several functionalities to the network edge in order to inherit advantage of the centralized computing and take advantage of the distributed computing [7]. A typical EC-RAN consists of four components: 1) centralized data cloud (CDC); 2) centralized control cloud (CCC); 3) distributed logical information processing cloud (DLIPC); and 4) fronthaul and backhaul (FnB) links. In this work, the communication links between CDC (and/or CCC) and DLIPC are named as backhaul, and the communication links between DLIPC and radio remote heads (RRHs) are named as fronthaul.

Leveraging the advantages of UAV technology and edge computing technology, we propose a novel architecture of EC-RAN with UAV-mounted FnB links to supplement the terrestrial infrastructure. Therefore, the sudden spike in traffic can be coped. In order to support the mobility of UAVs, the wireless FnB links will be utilized in the proposed EC-RAN [3]. Since the radio frequency (RF) spectrum is overcrowded, the implementation of RF-FnB links requires a complex protocol to guarantee the concurrent communication for the RF-FnB links and RF-access links. Based on the recent advancement of wireless optical communications [8]–[11], free space optical (FSO) communication becomes a promising FnB technique since FSO laser beams can deliver traffic over a long distance via license-free electromagnetic spectrum. Generally, the FSO laser beams operate beyond Terahertz spectrum (wavelength 750–1600 nm) [8]. Besides, the application of FSO-FnB links in the EC-RAN can benefit from the high probability of line-of-sight (LoS) propagation path since the UAVs cruise over the low (<1 km) to medium (1–10 km) altitude [1], [5].

While the current literature mainly deal with the overall investigation of UAV communications [1], design of UAV management framework [2], and verification of physical layer feasibility [3], our contributions are two folds: 1) we propose a novel EC-RAN architecture, where cache storage and light BBUs are mounted on the UAVs to supplement the terrestrial infrastructure via UAV-mounted FSO-FnB links; and 2) we focus on the issues related to networking technologies and CRM in the EC-RAN with UAV-mounted FSO-FnB links. Due to the unique challenges of the UAV-mounted FSO-FnB links, the issues related to networking technology and CRM are important for the successful deployment of the EC-RAN with UAV-mounted FSO-FnB links.

## II. An Architecture of Edge Computing Empowered Radio Access Networks

### A. Overall System Architecture

Fig. 1 illustrates the network architecture of EC-RAN with UAV-mounted FSO-FnB links. Specifically, the EC-RAN consists four layers: internet layer, cloud computing layer, edge computing layer and terminal layer. Moreover, the cloud computing layer and edge computing layer are used to provide the transparent wireless communication tunnels for the service providers (SPs) in the internet layer and the UEs in the terminal layer.

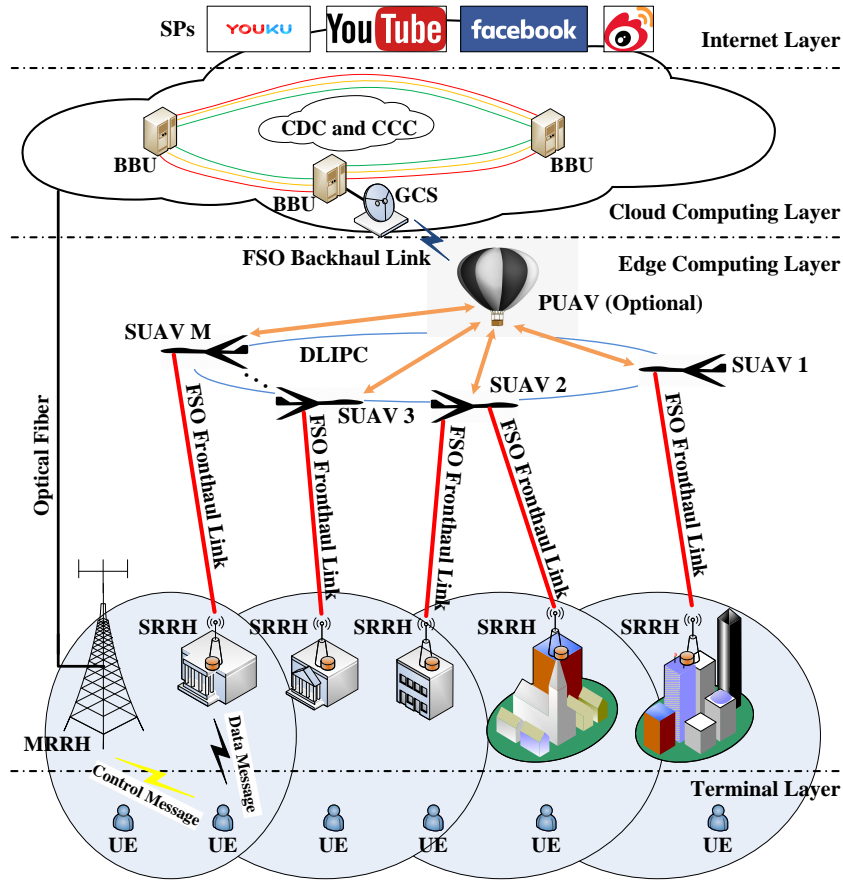

Fig. 1. An illustration of the EC-RAN with UAV-mounted FSO-FnB links.

In the cloud computing layer of the EC-RAN, the CCC and CDC consist of several base band units (BBUs) to perform the network wide control and data distribution functionalities, respectively. The CCC performs the big data analysis and optimizes the CRM decisions. For example, the CCC assigns a global unique identifier (GUI) to each content within the EC-RAN in order to avoid unnecessary duplication of the content in the EC-RAN. The CDC is mainly responsible for content storage and delivers the requested contents per the instruction of CCC.

As shown in Fig. 1, the edge computing layer consists of a single primary UAV (PUAV), multiple secondary UAV (SUAVs), a macro RRH (MRRH) and multiple small RRHs (SRRHs). The EC-RAN pushes some computing capabilities to the network edge via mounting a light BBU and caching media to each UAV. Moreover, the SUAVs in the DLIPC connect to the PUAV via FSO links. Note that the PUAV is quasi-static and optional to extend the coverage area of the UAV-mounted FSO-FnB links when the SUAVs are far away from the ground control station (GCS). If PUAV is employed, each SUAV updates the popular contents and feedbacks the statistics to cloud computing layer via the PUAV. Otherwise, the SUAVs establish FSO backhaul links with the GCS to perform these operations. The MRRH and SRRHs connect to the cloud computing layer via the optical fiber and UAV-mounted FSO-FnB links, respectively.

In the terminal layer of Fig. 1, the control messages of UEs are managed by the MRRH, and data messages of UEs are provided by the associated SRRHs. For example, MRRH has a larger coverage area than the SRRHs, and can act as the mobility management entity of UEs for the control messages, such as handover message and paging message. For further details on the mobility management, the interested reader is referred to [12, and references therein]. Each SRRH is also equipped with caching media to reduce the communication burden over the UAV-mounted FSO-FnB links. To provide indoor coverage, there is at least one SRRH placed on the rooftop of each building to connect to SUAVs, and the remaining SRRHs inside the building connect to the rooftop SRRH via optical fiber or coaxial cables.

## B. Hardware Implementations

Based on the state-of-the-art literature in [9]–[11], the hardware implementation is briefly discussed to keep the integrity of this article. Meanwhile, the interested readers are referred to [3] for the capital expenditure analysis of FSO based vertical fronthaul links.

*1) UAV Selection:* Based on different performance metrics, the classification of civilian UAVs is shown in Fig. 2. Compared with rotary-wing configuration, the fixed-wing UAVs can carry more payload and cover a larger area [1]. The electrical propulsion system for the UAVs has a small form factor and is reliable. The elimination of fossil fuel is another advantage of electrical UAVs, making the proposed EC-RAN economy-friendly and environment-friendly. Moreover, the evolution of the battery capacity and the usage of renewable energy and low-power FSO devices, the cruising endurance of the electrical UAVs can be significantly extended. Hence, the electrical UAV with fixed wings is selected as the configuration of SUAVs in the proposed EC-RAN with UAV-mounted FSO-FnB links. Since the PUAV needs to provide stable relaying between the GCS and SUAV, the medium-altitude-long-endurance balloon is chosen as the PUAV as shown in Fig. 1. The medium-altitude-long-endurance balloon, which is placed in quasi-stationary location, can generally provide a high payload for mounting the FSO devices, cache media and avionics.

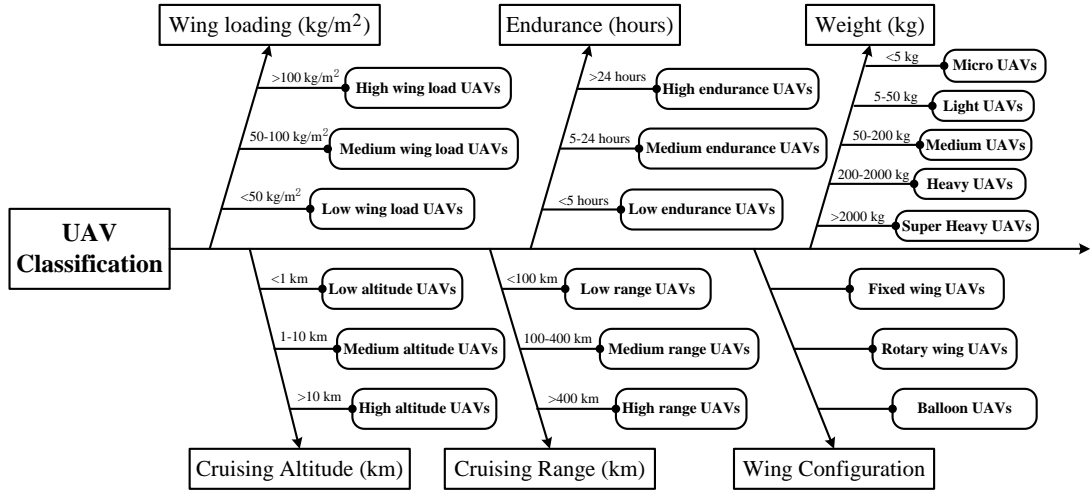

Fig. 2.   Civilian UAV classification based on the performance metrics.

*2) Retromodulator Based FSO Devices for the FnB Transceiver:* Considering the size, weight and power (SWAP) constraints of UAVs, the retromodulator based FSO device[1] becomes a good candidate as the major complexity of the transceiver pair is moved to the transmitter's or receiver's side. A typical retromodulator based FSO link consists of a conventional FSO transceiver (CFTRx) and a retromodulator based FSO transceiver (RFTRx) [9]. During the communication, the CFTRx illuminates the RFTRx with a modulated laser beam, from which the RFTRx can detect and demodulate the useful information. When the RFTRx needs to communicate with the CFTRx, the CFTRx illuminates the RFTRx with unmodulated laser beam. After obtaining the unmodulated laser beam, the RFTRx reflects back to the CFTRx with modulated laser beam. In this way, the two-way communication is implemented based on CFTRx and RFTRx. In addition, the CFTRx is mounted on the ground nodes (GCS and SRRHs) and high payload platform (PUAV), and the RFTRx is mounted on the payload constrained SUAVs. Since the RFTRx has low power consumption, small form factor and light weight, the major SWAP burden on the UAVs is offloaded to ground nodes and high payload platform [9]. The retromodulator based FSO devices can also provide a generally large field-of-view angle such that the laser beam can easily be aligned [9]. Moreover, the reflected laser via the retromodulator has the same direction of the unmodulated laser beam. Therefore, the complexity of beam alignment is only handled by SRRHs, PUAV and GCS.

---

[1]The size, weight and power consumption of a typical retromodulator based FSO device are, respectively, $55 \times 55 \times 50$ mm, 200 g, and 200 mW in DAZZLE project by Airbus Group Innovations.

*3) Position Tracking System for GCS, PUAV and SRRHs:* In order to communicate with the mobile UAVs, a global position system (GPS) and local position tracking system are required at GCS, PUAV and SRRHs. The local position tracking system consists of a time difference estimator, a steerable antenna and LoS alignment-and-tracking module [10]. Here, the steerable antenna can be implemented via a gimbal unit [11]. The GPS provides the initial coordinates and velocities of the RFTRxs. Then, the time difference estimator on the same entity of CFTRx obtains the time interval between two difference signals from the objective RFTRx. Based on the error correction Kalman prediction algorithm [10], the local position tracking system on the CFTRx can predict the position of the objective RFTRx via the LoS alignment-and-tracking module[2]. Finally, the gimbal assisted optical antenna can steer the direction of laser beam of the CFTRx towards the predicted position. In order to receive data from multiple laser beams simultaneously, each SUAV is equipped with multiple RFTRxs operating over different laser beam wavelengths with proper gap for inter-beam interference avoidance. During the detection and demodulation process, it is important to eliminate the cross talk between the laser beams coming from different SRRHs. In the coherent FSO communications, the cross talk elimination can be performed by frequency/polarization selectivity in optical domain. Such that the desired optical signal achieves optical gain while the interfering optical signals are suppressed [8].

*Remark 1:* Besides the CFTRx, each SRRH is also equipped with RF transceiver for UE-SRRH communications. With GPS and local position tracking system, the issue of LOS alignment-and-tracking can be well solved. The optical-booster amplifier and longer wavelength beam can be used to relieve the FSO signal degradation, which is mainly caused by the weather impairments, such as the absorption, scattering and atmosphere turbulence. Therefore, we abstract the FSO channel property and focus on the issues related to networking and CRM.

## III. Networking Technologies for EC-RAN With UAV-Mounted FSO-FnB Links

### A. Multiple Access Technology for SUAV-SRRH Links

To cope with the mobility and limited endurance of UAVs, a dynamic space multiple access scheme is a good candidate for the SUAV-SRRH links. From the perspective of CRM, the dynamic space multiple access corresponds to the connection resource management. Based on the hardware in Section II-B, the DLIPC has multiple SUAVs, each of which has limited number of established links. To exploit the two characteristics, the SRRHs can be classified into several groups based on the required contents. A specific group of SRRHs can search and connect of one of the SUAVs, which hold the required contents via orthogonal wavelengths. In case of no SUAV holding the required contents, some heuristic methods (e.g., closest UAV selection) can be used. Then, the required contents will be fetched via the GCS-PUAV-SUAV path.

---

[2]The commercial chipsets have already been released by some companies, e.g., ViaLight Communications.

### B. Multiplexing Technology for PUAV-SUAV Link

Using the gimbal aided optical antenna, a time division beam multiplexing scheme can be used to establish the point-to-multipoint communications for the PUAV-SUAV link. Based on the position tracking system, the PUAV are able to acquire and track the position of the objective SUAVs. Therefore, the contents requested by SUAVs can be obtained via the relaying of PUAV.

### C. Mobility Handover Technology of SRRHs

The mobility of SUAVs results in frequent handover operations, which lead to severe signaling overhand, transmission disruption and increased latency. Therefore, a suitable triggering mechanism is required to minimize the negative effects of the frequent handover of SRRHs to SUAVs. Compared with the RF counterpart, the reference signal strength (RSS) of FSO signals fluctuates less in terms of multipath fading and has negligible doppler effect [11]. Hence, the RSS is selected as the criterion to trigger handover.

As the received RSS at a SRRH drops below a predefined threshold, the SRRH starts to search next available SUAVs. The procedure of RSS based handover is shown in Fig. 3. Initially, the SRRH sends a handover request via the beacon laser beam to the nearby SUAVs when SRRH identifies that RSS from original UAV is below the threshold. Then, each receiving SUAV replies to the requesting SRRH with a beacon laser beam if it is available to accommodate the requesting SRRH. Finally, the SRRH selects the SUAV that provides beacon laser beam with the highest RSS. As a result, the requesting SRRH terminates the connection with the original SUAV and establishes a connection with the selected SUAV. The main advantage of RSS triggered handover is the low implementation complexity since the SRRHs performing the handover only needs to compute the received RSS power. In the RRH based handover, the predefined threshold needs to be chosen in order to avoid frequent handover and the resultant deleterious effects.

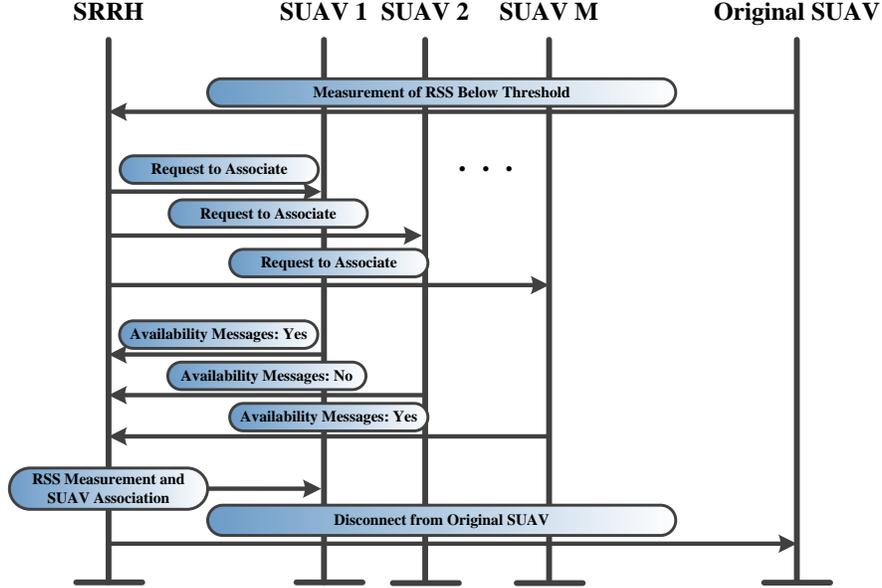

Fig. 3.   RSS based handover procedure for the SUAV-SRRH links.

## D.  Use Case Study

As shown in Fig. 4(a), we investigate a downlink use case where the SRRHs are randomly distributed over a 300-meter long road, and the PUAV are SUAVs are cruising over an altitude of 300 meters following the channel setting [3], [9]. The PUAV and GCS are deployed 2500-meter and 3000-meter far from the center of the trajectories, respectively. The trajectory is an ellipse with parameter $A$ as $\frac{(x-150)^2}{9A^2} + \frac{(y-50)^2}{A^2} = 1$. The detailed simulation parameters are shown in Fig. 4(b).

Fig. 5(a) shows that the system end-to-end throughput of the EC-RAN with UAV-mounted FSO-FnB links decreases with increasing weather attenuation caused by fog, cloud and rain [3]. On the other hand, the UAV-mounted RF-FnB links are not sensitive to the weather conditions. When the weather attenuation is larger than 57 dBm/km, the PUAV is required to perform as a relay to compensate for the weather attenuation. We also simulate performance of the proposed network architecture with PUAV, and the system end-to-end throughput achieved by the UAV-mounted FSO-FnB links still outperforms that of the UAV-mounted RF-FnB links (c.f. Fig. 5(a)). As shown in Fig. 5(b), the throughput of the FSO-FnB links without PUAV firstly increases with the value of $A$, and decreases when the value of $A$ is larger than 45 m. This observation might motivate the operator to optimize the trajectory of SUAVs to exploit the performance gain of

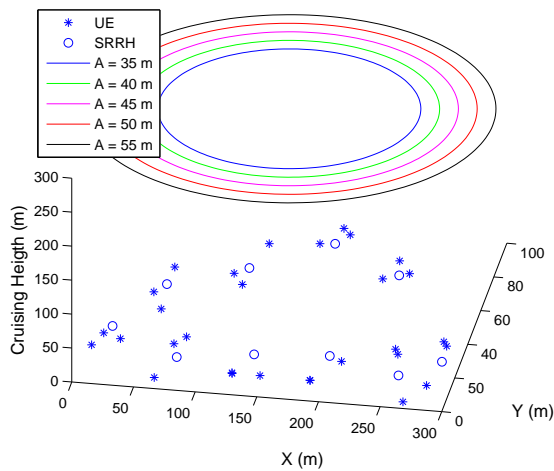

(a) Trajectory of SUAVs.

| Parameters | Values |
|---|---|
| # of SUAVs | 5 |
| # of SRRHs | 10 |
| # of UEs of each SRRH | 3 |
| # of points on trajectory | 180 |
| Tx power of SRRH, PUAV, GCS and SUAV | 200 mW |
| Noise power | 1e-10 mW |
| Aperture of SRRH, PUAV and GCS | 5 cm |
| Aperture of Retromodulator | 1 cm |
| FSO wavelength | 850 nm |
| Speed of wind | 20 m/sec |
| RF antenna gain of SRRH | 5 dBi |

(b) System parameter setting.

Fig. 4. System parameter setting and trajectory of SUAVs.

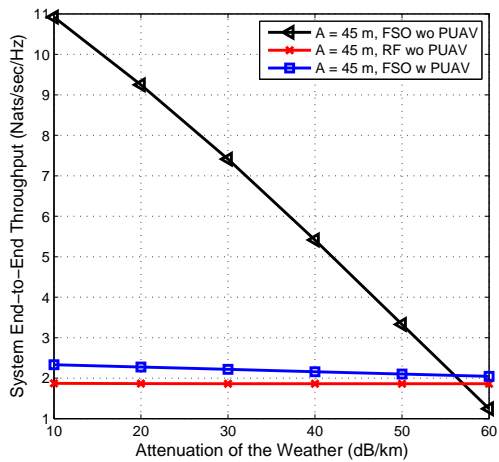

(a) Weather attenuation v.s. throughput.

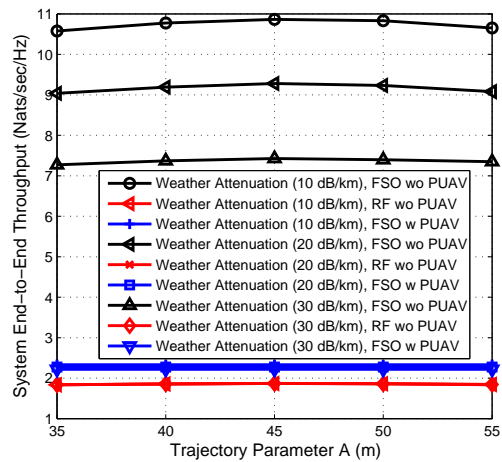

(b) Trajectory parameter v.s. throughput.

Fig. 5. System end-to-end throughput.

the SUAVs. Since the GCS-PUAV link is the bottleneck, the throughput of the FSO-FnB links with PUAV and RF-FnB links without PUAV is not sensitive to the value of $A$. Therefore, an intelligent caching strategy is required to reduce the burden of backhaul link between PUAV-GCS link.

## IV. CRM ISSUES AND RECENT RESEARCH PROGRESSES

In the proposed EC-RAN architecture with UAV-mounted FSO-FnB links, the term resource is generalized from radio resource (e.g., frequency bands, time slots, code chips and transmission power) to include computation resource (e.g., BBUs), storage resource (e.g., cache), mobility resource (e.g., trajectory) and connectivity resource (e.g., FSO-FnB links). Since the FSO technology is used to support the FnB links, the characteristics of the FSO communication and the mobility of the UAVs make the CRM issues in trajectory plan and caching content placement challenging. Hereinafter, we mainly discuss the algorithm design related to trajectory plan and caching content placement.

### A. Trajectory Plan of SUAVs in 3D Space

In the context of the EC-RAN, the high mobility of SUAVs increases the coverage area of the DLIPC. However, the high mobility also introduces significant differences between the vertical channels (SUAV-SRRH channels and PUAV-SUAV channels) and horizontal channels (SRRH-UE channels). The limited endurance also confines the cruising distance of each UAV. Hence, the trajectory plan of UAVs in the DLIPC is of vital importance to fulfill the requirements for the UAV-mounted FnB links.

The trajectory optimization of UAVs has been investigated for various applications [4], [5]. Due to the high mobility of UAVs, it becomes challenging to establish communication links for vertical channels while considering the communication quality of service. For example, the link on a vertical channel varies over time, which leads to severe delay jitter. Thus, the maintenance of satisfactory communication becomes a key issue in the trajectory plan of UAVs. In [4], the authors proposed a centralized dynamic trajectory control algorithm to improve the success probability of a communication link establishment for the buffer-limited UAVs. When the buffer of a UAV overflows, a congestion happens; therefore, a new communication link of the UAV cannot be established. To reduce the congestion at the UAV, a GCS can move the trajectory center of neighboring UAVs towards the congested UAV. Moreover, the GCS can configure the trajectory radius of the UAVs to provide sufficient communication links to the objective area. Besides, the endurance constraint of UAVs makes the energy efficiency optimization another crucial topic in the trajectory plan of UAVs. The authors in [5] pointed out that the propulsion system of a UAV dominates the energy usage in the UAV communications, and theoretically proved that the throughput maximization and energy minimization solutions to be energy inefficient. Therefore, they investigated the energy efficiency maximization of a point-to-point UAV communication system and proposed an energy efficient trajectory optimization algorithm.

The UAV communication links operate on RF spectrum for vertical channels in the state-of-the-art work [4], [5]. However, the RF spectrum suffers from a bandwidth scarcity and limited capacity, which requires efficient interference management. Due to the high link capacity and the interference-free characteristic, the combination of the FSO communications with the UAV systems becomes an appealing alternate for the UAV-mounted FnB links in the EC-RAN. To avoid the blockage of the clouds for the UAV-mounted FnB links, the trajectory plan of UAVs in the DLIPC is extended to three dimensions (3D), i.e., longitude, latitude and altitude. Specifically, the cruising altitude of a blocked UAV can be reduced by flying below the cloud. Motivated by [4], [5], the energy efficient 3D trajectory plan of UAVs in the DLIPC needs to be developed based on the UAV propulsion energy consumption model [5] via tuning the velocity, acceleration and altitude of the UAVs in the DLIPC of the EC-RAN. Besides, the trajectory of the UAVs influences the performance of SUAV-SRRH links, such as received signal strength and delay with respect to receiving the cached contents in the UAV. Hence, trajectory plan should be jointly optimized with cache content placement and UAV-mounted FnB link association.

### B. Intelligent Caching Strategy of UAVs

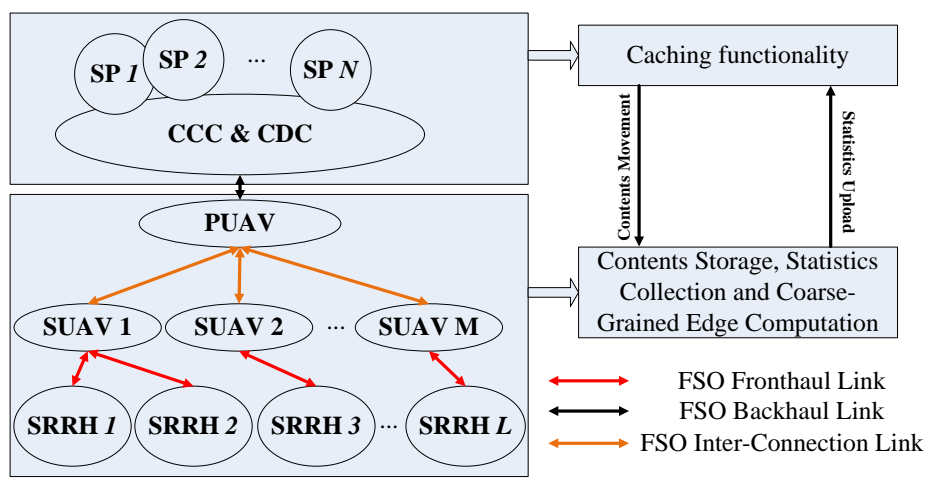

Fig. 6. The FCaaS framework in the EC-RAN.

An intelligent caching strategy enables the analysis of content requirement and proactively prefetches the popular contents in order to reduce the redundant transmission according to the "Power Law" effect of content popularity[3] [6], [13], [14]. As shown in Fig. 6, a caching virtualization framework, named as Flying Cache as a Service (FCaaS), is proposed in the proposed EC-RAN. Within the FCaaS framework in the EC-RAN, the CCC, SUAVs, SRRHs and UEs maintain the same GUI table of the popular contents. The index of the GUI table of each node equals to one if the content is cached in the node in the FCaaS framework; otherwise, it equals to zero. The computation-massive caching functions are performed by the CCC after obtaining the global network knowledge, such as channel state information, UAV-mounted FnB link availability and etc. Before submitting the local network knowledge to the CCC, the UAV-mounted BBUs perform preprocessing, e.g., coarse-grained content popularity classification, to reduce the burden of the CCC. The FCaaS framework enables that the CCC in the EC-RAN to exploit benefits of the centralized fine-grained computing resource and distributed coarse-grained computing resource for caching functionalities. The caching instances can be created and configured adaptively according to the network demands with great flexibility. In addition, the popular content update becomes real-time compared with the terrestrial counterpart that needs to update popular contents during the off-peak hours.

In the proposed EC-RAN, the coverage area of the DLIPC is larger than that of a single SRRH. This characteristic leads to diverse demands on the cached contents of the UAVs from the terrestrial SRRHs, which translates into high storage requirements on the cache of the UAVs. However, the available payload of a UAV is limited due to the communication and avionic devices. Therefore, a cooperative caching strategy is required to fully exploit the limited caching resource and the edge-computing resource [13]. Leveraging the GUI tables in FCaaS framework, the cooperative caching strategy can be developed based on the mobility of UAVs and the popularity of different contents. For example, in order to reduce the content delivery for the SUAV-SRRH links, replica of the most popular contents require to be cached at each SRRH, which can be implemented based on setting the indices of the same contents in the GUI table of each SRRH as one. The duplication of contents may reduce the content hit ratio and increase the burden on the UAV-mounted FSO-FnB links. Since the popularity of contents are highly region oriented [14]: residential region for entertainment contents and technological business district for work-related contents, the CCC can use the fine-grained computing resource to figure out the type of a specific region and prefetches the required contents to their suitable region. Moreover, each UAV in the DLIPC can prefetch the popular contents that are not cached in the terrestrial SRRHs via performing the XOR operation of the popular content table with the GUI tables of the covered SRRHs. Besides, the prefetching operations also need to consider the mobility of UAVs in the cooperative caching strategy. For example, a content prefetching request is sent from SUAV 1 to PUAV at the time instance $t_0$. The requested content arrives at the PUAV at time instance $t_0 + \Delta t$ from the CDC. However, the SRRH, which was connected to the SUAV 1 at the time instance $t_0$, connects to the SUAV 2 at time instance $t_0 + \Delta t$. Thus, PUAV should send the requested content to the SUAV 2 instead of SUAV 1.

---

[3]The most popular contents account for small amount of overall contents in the network. For example, the top 10% of the contents account for the 80% subscription in Youtube.

## V. CONCLUSION AND FUTURE WORK

We proposed an architecture of the EC-RAN with UAV-mounted FSO-FnB links. The proposed architecture is envisioned to supplement the terrestrial infrastructure. Based on the state-of-the-art of hardware implementation, we outlined the issues related to networking technology and CRM. Moreover, we provided several numerical evaluation to show the performance improvement of the FSO-FnB links over the RF-FnB links.

In order to gain theoretical insights of the proposed EC-RAN, we envisage that the framework of Lyapunov optimization can be used to articulate the CRM problems since it considers instantaneous network states allocation to optimize the long-term performance metric subject to long-term network dynamic metrics [15]. In the Lyapunov optimization framework, the network dynamic metrics are abstracted as several virtual queues. Therefore, the long-term network dynamic metrics are satisfied by guaranteeing the virtual queues stable. The major challenge in the Lyapunov optimization framework lies in the transformation of the formulated problem into Lyapunov-drift-plus-penalty function [15].